\documentclass[aps, prl, reprint, showpacs, superscriptaddress]{revtex4-1}
\usepackage{amsmath, amsfonts, amssymb, bm}
\usepackage{color, graphicx}
\usepackage[colorlinks,citecolor=blue, urlcolor=blue,
  linkcolor=blue]{hyperref}
\begin{document}

\title{Competing phases of the Hubbard model on a triangular lattice --
  insights from the entropy} 

\author{Gang Li}
\affiliation{\mbox{Institut f\"ur Theoretische Physik und Astrophysik,
  Universit\"at W\"urzburg, 97074 W\"urzburg, Germany} }

\author{Andrey E. Antipov}
\affiliation{\mbox{Max Planck Institute for Chemical Physics of
    Solids, 01187 Dresden, Germany}}
\affiliation{\mbox{Max Planck Institute for the Physics of Complex
    Systems, 01187 Dresden, Germany}} 

\author{Alexey N. Rubtsov}
\affiliation{ \mbox{Department of Physics, Moscow State University, 119992
    Moscow, Russia} }
   
\author{Stefan Kirchner}
\affiliation{\mbox{Max Planck Institute for Chemical Physics of
    Solids, 01187 Dresden, Germany}}
\affiliation{\mbox{Max Planck Institute for the Physics of Complex
    Systems, 01187 Dresden, Germany}} 

\author{Werner Hanke}
\affiliation{\mbox{Institut f\"ur Theoretische Physik und Astrophysik,
  Universit\"at W\"urzburg, 97074 W\"urzburg, Germany} }

\begin{abstract}
  Based on the ladder dual-fermion approach, we present a comprehensive
  study of the phases of the isotropic Hubbard model 
   on the triangular lattice. We find a rich phase
  diagram containing most of the phases that have already been
  experimentally observed in systems where the interplay between
  geometric frustration and electronic correlations is important:
  paramagnetic metal, paramagnetic insulator, Mott-insulator with
  $120^{\circ}$ antiferromagnetic and a non-magnetic insulating state,
  {\it i.e.} possibly a spin liquid state. 
  This establishes that the Hubbard model on frustrated lattices can
  serve as a minimal model to address the intricate interplay of
  frustration and correlation. 
  We also show that entropic considerations can be successfully used
  for understanding many striking features of the triangular systems,
  such as the large thermopower found in Na$_{x}$CoO$_{2}\cdot$$y$H$_{2}$O.
\end{abstract}


\maketitle

{\it Introduction.}
Spin models with frustrated interactions, caused either  by geometric
frustration of the underlying lattice or due to competing
interactions, can differ significantly in their physical properties
from their non-frustrated counterparts. In a frustrated system, a spin
cannot find a configuration that simultaneously minimizes all its
interactions with its neighbors. As a result, spin frustration may completely 
suppress the long-range magnetic ordering in a system, resulting in a
non-magnetic insulating ground state that does not break any symmetry,
{\it i.e.} a spin liquid (SL) state~\cite{2010Natur.464..199B}. 
Real magnetic materials are typically characterized by frustrated
interactions which give rise to  rich phase diagrams.
A SL state, may  have been
observed {\it e.g.}  in the organic salts
$\kappa$-(BEDT-TTF)$_{2}$Cu$_{2}$(CN)$_{3}$~\cite{PhysRevLett.91.107001}, 
where the frustration among the quantum spins is
due to the underlying triangular lattice.
In contrast, the organic charge transfer salt
$\kappa$-(BEDT-TTF)$_{2}$Cu[N(CN)$_{2}$]Cl displays anti-ferromagnetic
(AF) long-range order at low
temperatures~\cite{PhysRevLett.91.016401}. 
The $\kappa$-(BEDT-TTF)$_{2}$X family and the layered cobaltate
Na$_{x}$CoO$_{2}\cdot$$y$H$_{2}$ are triangular systems that give rise to
superconductivity (SC)~\cite{2003Nature.422.53, 2003Nature.424.527,
  PhysRevLett.92.247001}. 
It is an open question if the emergence of SC in these systems can be
connected to the SC phases observed in the high-T$_c$ cuprates.
Moreover, these systems also display Fermi-liquid behavior and give
rise to Mott insulating behavior, as well as to transitions between
a SL and antiferromagnet (AFM)~\cite{PhysRevLett.91.107001,
  PhysRevB.77.104413}. 
We also note that an unusually large thermopower is found in the cobaltate
Na$_{x}$CoO$_{2}\cdot$$y$H$_{2}$~\cite{Minhyea2006}.

The existence of electronic phases on frustrated lattices indicates
the necessity to consider itinerant electron models as effective low
energy models for these systems that allow for the interplay between
electron correlation and geometric frustration. 
All materials mentioned above have an effective triangular structure
and are characterized by strong electronic correlations.
This raises the important question if Hubbard-type models on the
triangular lattice  constitute a model class that can encompass
the experimentally observed phases as possible ground states. 

\begin{figure}[htbp]
\centering
\includegraphics[width=\linewidth]{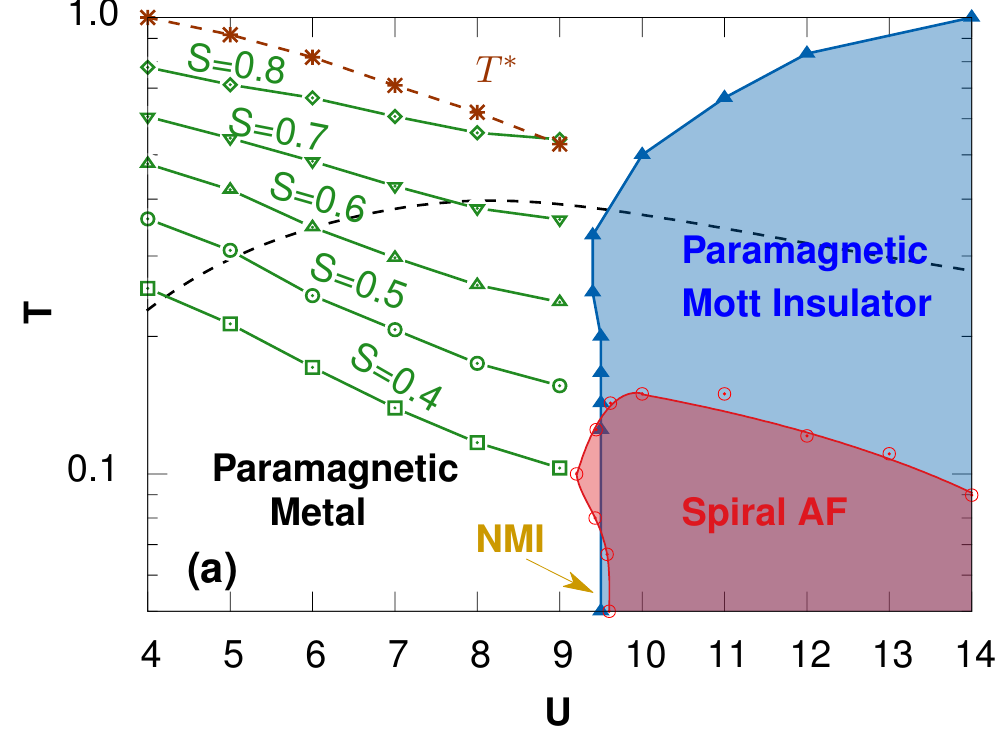}
\includegraphics[width=\linewidth]{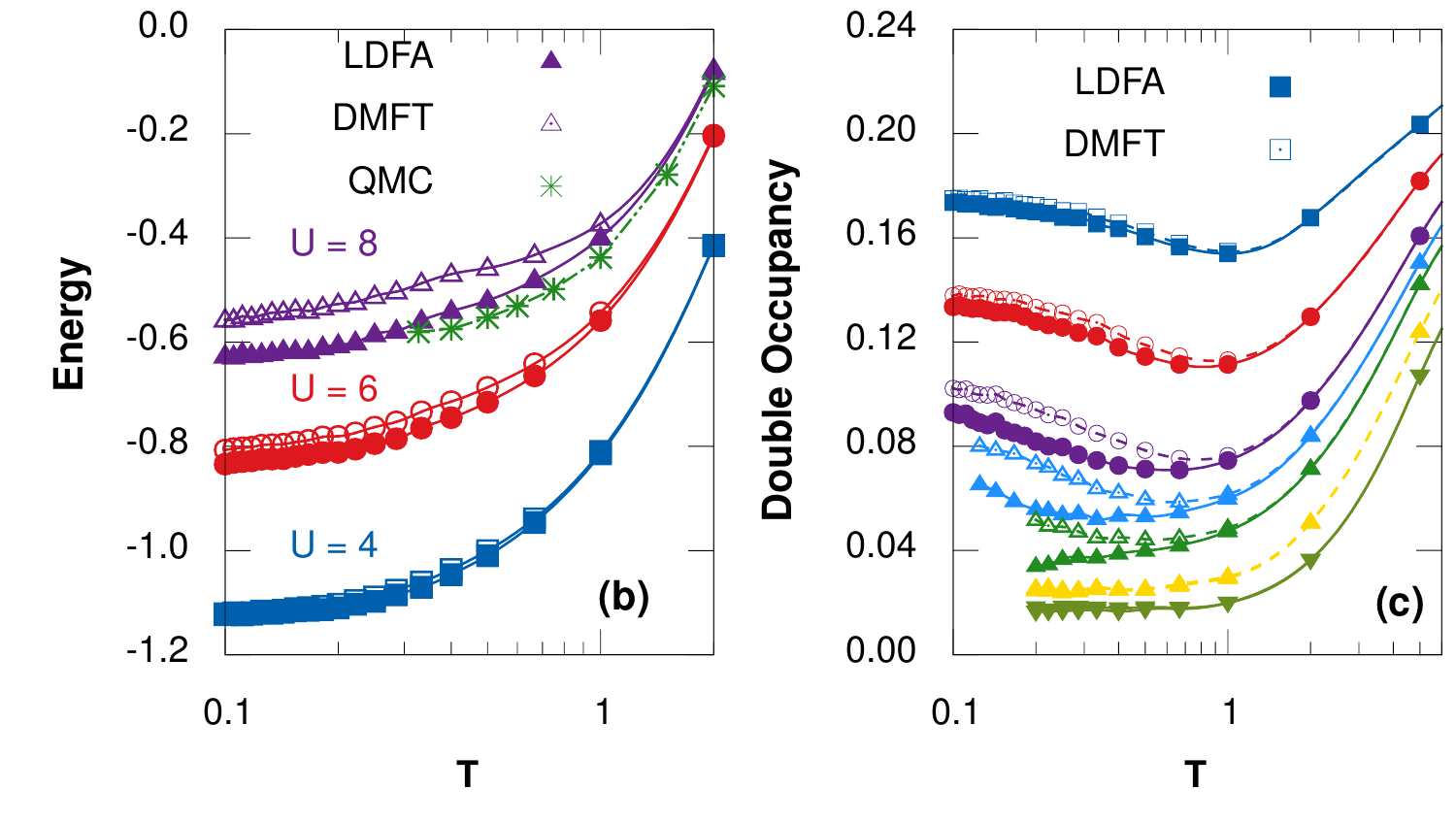}%
\caption{(a). $T$-$U$ phase diagram of the half-filled ITH 
  model. The monotonic decrease of the constant-entropy
  curve reveals the possibility of adiabatic cooling in this
  frustrated system. 
$T^{*}$ is the corresponding temperature at which the double occupancy
  in Fig.~\ref{PhaseDiagram-Half}(c) is  minimized. 
  The NMI phases are found at lower temperature region ($T\le0.08$)
  and for $U$ slightly larger than the band width ($U\in[9.4,
    9.55]$).
  See text for more details.
  (b). Comparison of the total energy $E_{U}(T)$ calculated from the
  DMFT (empty symbols) and the LDFA (solid symbols)
  calculations. Their difference results  from the non-local 
  correlations. The QMC results for $U=8$ are extracted from
  Ref.~\cite{PhysRevB.74.085117}.
  (c). The double occupancy as a function of temperature for different
  interactions (From top to bottom: $U=4,
  6, 8, 9, 10, 12, 14$, respectively.)}
\label{PhaseDiagram-Half}
\end{figure}

\begin{figure}[htbp]
\centering
\includegraphics[width=\linewidth]{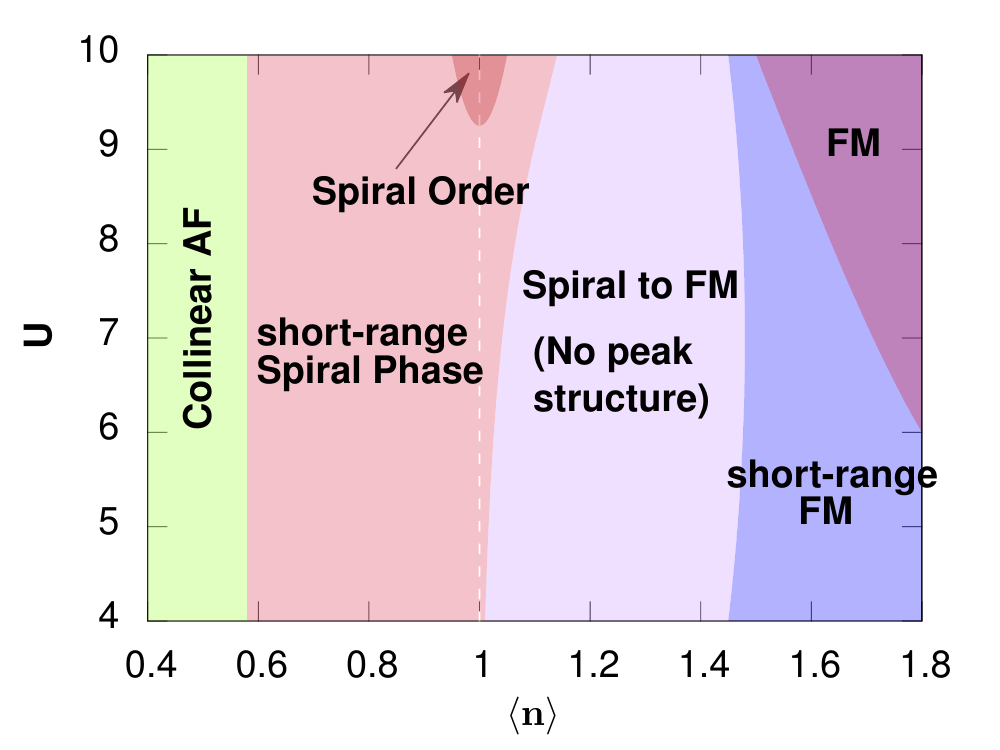}
\caption{Magnetic phase diagram of the doped triangular Hubbard model
  as a function of the Coulomb strength at $T=0.1$. 
  Two stable magnetic phases are found, {\it i.e.} a spiral($120^{\circ}$)-AF
  (anti-ferromagnetic) and a ferromagnetic (FM) phase. Short-ranged 
  $120^{\circ}$-AF and FM states are found in the vicinity of
  long-range $120^{\circ}$-AF and FM phases. }
\label{PhaseDiagram-Away}
\end{figure}

In the present letter, we study the finite-temperature thermodynamic,
electronic, and magnetic properties of the doped isotropic triangular
Hubbard (ITH) model. The resulting phase diagram is
summarized in Figs.~(\ref{PhaseDiagram-Half})(a) (at half filling) and
(\ref{PhaseDiagram-Away}) (away from half filling). 
One standard approach to strongly correlated
electronic systems is based on the dynamical mean-field theory
(DMFT)~\cite{RevModPhys.68.13}.
The inclusion of the effects of geometric
frustration, however,  requires an extension beyond the single-site problem.
Faithfully capturing the competition between strong correlations and
geometric frustration across the phase diagram is not only essential
but also very challenging.
Our study is based on a powerful technique, 
{\it i.e.} the ladder dual-fermion approach
(LDFA)~\cite{PhysRevLett.102.206401}. 

The dual fermion approach~\cite{PhysRevB.79.045133} is a
non-local extension of the DMFT. It introduces a set of dual variables
with an effective interaction formed by the reducible two-particle
vertex of the interacting fermions as obtained from the DMFT.
An interaction expansion over the dual variables yields systematic
non-local corrections to the standard DMFT approach. 
Studies have shown that already the inclusion of the first two
lowest-order dual diagrams produces good agreement with numerically
exact results~\cite{PhysRevB.43.8044}. 
Including higher order diagrams further improves the
agreement~\cite{PhysRevLett.102.206401}.
In what follows, we take the hopping parameter $t$ of the ITH as our
unit of energy, {\it i.e.} $t=1$.

Our choice of method for addressing the ITH is motivated by the
advantages it offers over possible alternatives:
(1) the calculations are not restricted to  finite-size clusters, {\it
  i.e.} the  thermodynamic limit and nonlocality are naturally
incorporated, 
(2) although a quantum Monte Carlo simulation is
involved~\cite{PhysRevB.72.035122}, the calculations are free of the
``minus"-sign problem (everywhere in the phase diagram), 
(3) both spin and charge collective excitations are accessible which
allows us to establish the nature of various insulating and magnetic
phases.  
In what follows, we take the hopping parameter $t$ of the ITH as our
unit of energy, {\it i.e.} $t=1$.

{\it Results.}
The magnetic phase diagram of the half-filled ITH model within our approach is
shown in Fig.~\ref{PhaseDiagram-Half}(a). 
In analogy to its counterpart on a square lattice~\cite{PhysRevB.83.085102,
  RevModPhys.68.13}, 
three major phases are found for the ITH: a paramagnetic metal at smaller
interaction $U$, a paramagnetic insulator at larger $U$ and higher
temperature $T$, and 
a Mott insulator with $120^{\circ}$-AF order at larger $U$ and lower $T$.

As our finite-temperature LDFA calculations are carried out on
discrete Matsubara frequencies,
the exact location of the metal-insulator transition (MIT) boundary is
hard to determine.
In this work, we identify the first-order transition line from the
flatness of $G^{LDFA}_{ii}(i\omega)$ at the two lowest Matsubara
frequencies~\cite{Roman2012}.
The magnetic transition boundary is extracted from the extrapolation
of the inverse spin susceptibility $\chi^{spin,
  -1}_{\Omega_{m}=0}(\mathbf{Q}=\mathbf{K})$ as a function of
temperature $T$ and interaction $U$.
(see examples in the supplemental material), where $\mathbf{K}$ is the
magnetic wave vector for the 120$^{\circ}$-AF order.

The influence of geometrical frustration on the magnetic properties is
clearly visible, see  Fig.~\ref{PhaseDiagram-Half}(a): 
the onset temperature of the $120^{\circ}$-AF (also called spiral-AF)
state is significantly smaller than its counterpart for unfrustrated
lattices (for the square lattice, this corresponds to the black dashed
line of Fig.~\ref{PhaseDiagram-Half}(a)~\cite{PhysRevB.83.085102}).  
While for  the Hubbard model on a square lattice, AF order exists for
any non-vanishing value of $U$, the geometrical frustration of the
triangular system suppresses the long-range magnetic order at smaller
$U$ down to zero temperature, resulting in a non-magnetic metal below
$U\sim9.55$. 
Here, the onset temperature of the $120^{\circ}$-AF order (which is
strictly speaking absent in 2D~\cite{PhysRevLett.17.1133}), obtained
from our calculations, can be viewed as the upper limit of the N\'eel
temperature for a triangular system with a larger dimension, e.g. by
coupling the triangular lattice into multilayers.

A very interesting effect of  geometric frustration is that it
pushes the magnetic transition boundary
towards higher $U$ values to a value close to $U_{c}^{MIT}$, opening 
up the possibility of a ground state characterized by a nonzero charge
gap and the absence of any long-range magnetic order(see
Fig.~\ref{PhaseDiagram-Half}(a)). 
This is indeed what we find within the LDFA:
there is a non-magnetic
{\it Mott} insulating (NMI) phase in the interval
$U\in[9.4, 9.55]$ for $T\le0.08$~\cite{PhysRevB.83.115126}.  
The appearance of the NMI phase is highlighted by the crossing of magnetic
and insulating phase boundaries in Fig.~\ref{PhaseDiagram-Half}(a).
We find for the critical value $U_{c}^{MIT}=9.4$ at $T= 0.2$ 
and $U_{c}^{MIT}$ hardly changes upon further decreasing $T$.
The magnetic transition from the paramagnetic to
the $120^{\circ}$-AF state, on the other hand, shows a clear temperature
dependence, {\it i.e.} $U^{Spin}_{c}$ becomes larger with lowering the temperature
for $T\le0.11$.
At the lowest temperature studied ($T_{\mbox{\tiny low}}=0.05$),
long-range order  is established at $U=9.55$, leaving the phase with 
$U\in[9.4, 9.55]$ to be a NMI. Note that this conclusion is free of
any finite-size effects but the phase boundaries in
Fig.~\ref{PhaseDiagram-Half}(a) are subject to extrapolations (see
supplemental material).  
The NMI is a natural candidate for the SL phase at zero-temperature 
and our results are in line with experimental findings for
$\kappa$-(BEDT-TTF)$_{2}$Cu$_{2}$(CN)$_{3}$~\cite{PhysRevLett.91.107001}.
The bulk spin susceptibility of
$\kappa\mbox{-(BEDT-TTF)}_{2}\mbox{Cu}_{2}\mbox{(CN)}_{3}$ shows no 
sign of long-range antiferromagnetic order at significantly
lower temperature as compared to the Heisenberg exchange estimated
theoretically from the high-temperature series
expansion~\cite{PhysRevB.71.134422}.

In order to assess the importance of non-local correlations, we analyze the
 total energy $E_{U}(T)$ for the ITH within DMFT-only and  LDFA calculations,
see Fig.~\ref{PhaseDiagram-Half}(b).
The difference between the DMFT and the LDFA solutions at fixed $T$ 
signals the importance of non-local fluctuations, which is particularly large
when $U$ is close to $U^{MIT}_{c}$. 
Note that $E_{U}(T)$ from DMFT is never smaller than the one obtained
from  the LDFA. 
At $U=8$, the LDFA results show a good agreement with a determinant QMC 
simulation~\cite{PhysRevB.74.085117} for all temperatures sampled. 
Thus, our LDFA calculations systematically incorporate non-locality
into DMFT and  correctly resolve the total energy at low
temperatures. 
At low temperatures the DMFT results clearly differ from the
determinant QMC results.  
Thus, Fig.~\ref{PhaseDiagram-Half}(b) demonstrates that {\it non-local
correlations} are significant and cannot be neglected.
These correlations stabilize the {\it Mott} insulating
phase~\cite{0295-5075-84-3-37009, PhysRevLett.101.186403,
  PhysRevB.78.205117} and suppress spin
correlations~\cite{PhysRevB.78.195105, PhysRevB.77.195105}, 
thereby favoring the NMI phase at lower temperatures.

Another important observation that can be read off from
Fig.~\ref{PhaseDiagram-Half}(a) is, 
that the constant-entropy curves monotonically decrease with
increasing $U$. 
The entropy is obtained from the total energy by integration,
{\it i.e.} $S_{U}(T, n) = S(0, n) + E_{U}(T, n)/T - \int_{T}^{\infty}
E_{U}(T^{\prime})/T^{\prime, 2} d T^{\prime}$, with $S(0, n)=-n\ln
n/2-(2-n)\ln(1-n/2)$, where $n$ is the average filling.  
For numerical stability, we fitted the total energy $E_{U}(T)$
with an exponential function~\cite{PhysRevB.63.125116} and also
cross-checked with a two-segment fitting
scheme~\cite{PhysRevB.80.140505, PhysRevB.74.085117}, which resulted in
the same qualitative behavior of $S_{U}(T)$. 
The entropy is related to the number of doubly occupied sites $D$ through a
thermodynamic relation, i.e. $\partial S/\partial U=-\partial
D/\partial T$.  
As shown in Fig.~\ref{PhaseDiagram-Half}(c), the double occupancy $D$ has
a negative slope for temperatures below the critical value
$T^{*}$, thus, resulting in an increase of $S$ as $U$ increases for
fixed $T$ (Fig.~\ref{PhaseDiagram-Half}(a)).  
This has important ramification for cold-atom studies in that it
allows for adiabatic cooling of the system by increasing the 
Coulomb interaction~\cite{PhysRevLett.95.056401, Wolf26042011}.

The relation above can be rewritten as
$C/T(\partial T/\partial U)_{S}=(\partial D/\partial T)_{T}$ where $C$
denotes the specific heat.
This immediately implies that the isentropic ({\it i.e.} keeping the
entropy constrant) increase of $U$ results in a decrease of $T$
for $T\le T^{*}$. 
As one can see from Fig.~\ref{PhaseDiagram-Half}(a), the decreasing
behavior of the entropy ends at $T^{*}$.  This is due to
the opening of the charge gap, resulting in a saturation of $D$  at
lower temperatures~\cite{PhysRevLett.95.056401}. The 
$T^{*}$ curve (by extrapolation) ends at the boundary of the
insulating phase in Fig.~\ref{PhaseDiagram-Half}(a). 
We note that the insulating phase boundary in
Fig.~\ref{PhaseDiagram-Half}(a) is determined from the opening of the 
charge gap, which, as expected, coincides with the saturation of $D$
(see Fig.~\ref{PhaseDiagram-Half}(c)) in our LDFA
calculations.

We now turn to the doping dependence away from half-filling.
Fig.~\ref{PhaseDiagram-Away} displays the magnetic phase diagram of the
ITH model as a function of doping  and $U$.
As discussed in Fig.~\ref{PhaseDiagram-Half}(a), the $120^{\circ}$-AF
phase is stabilized at $T=0.1$ for $U > 9.5$ at half filling ({\it i.e.}
$\langle n\rangle=1$).  
This phase extends slightly away from the half-filling case, as
shown in  Fig.~\ref{PhaseDiagram-Away}. 
The stability of the $120^{\circ}$-AF phase is characterized by the
divergence of $\chi_{\Omega_{m}=0}^{spin}(\mathbf{Q})$ at
$\mathbf{Q}=\mathbf{K}$. 
By doping the system away from half-filling, the spiral-AF phase is quickly
destroyed, resulting in a very narrow phase region (see 
Fig.~\ref{PhaseDiagram-Away}). 

\begin{figure}[htbp]
\centering
\includegraphics[width=\linewidth]{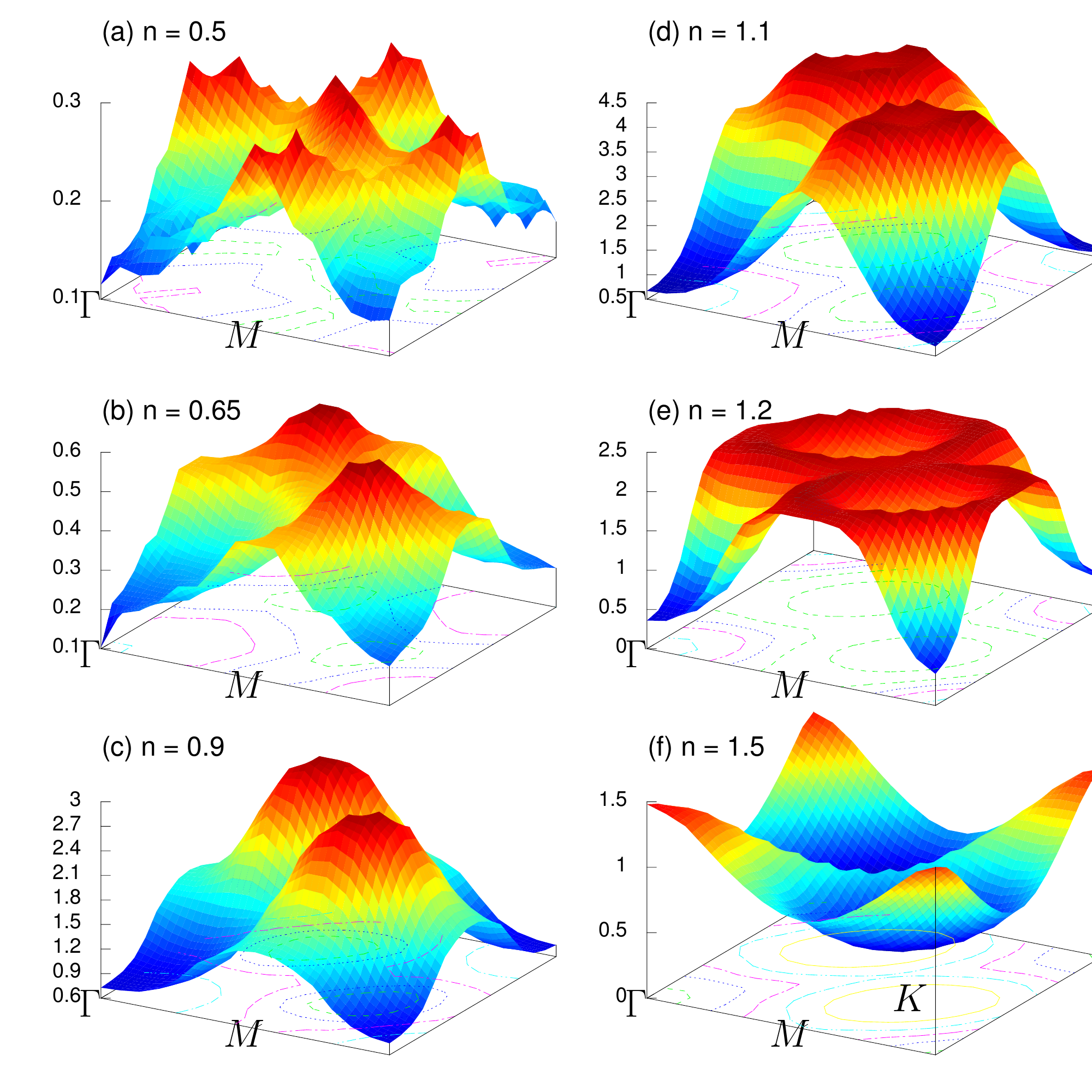}
\caption{Spin susceptibilities
  $\chi_{\Omega_{m}=0}^{spin}(\mathbf{Q})$ extracted from the LDFA
  scheme for six different doping levels and $\beta= 10$, $U = W$ 
with $W$ the bandwidth. $\mathbf{\Gamma}$, $\mathbf{M}$ and
$\mathbf{K}$ are the high-symmetry points of the 1st BZ of the
  triangular lattice. The 1st BZ is replotted here equivalently as a
  square.}
\label{Sus-Peak}
\end{figure}
At the electron doped side, the destruction of the $120^{\circ}$-AFM is
characterized by the removal of the spin susceptibility peak at
$\mathbf{K}$. 
As shown in Fig.~\ref{Sus-Peak}(d), at $\langle n\rangle =
1.1$, the spin susceptibility peak at $\mathbf{Q}=\mathbf{K}$ starts
to disappear. 
Further increasing electron doping completely removes the peak, giving
rise to a flat structure of $\chi_{\Omega_{m}=0}^{spin}(\mathbf{Q})$
at $\mathbf{Q}$ around $\mathbf{K}$ (Fig.~\ref{Sus-Peak}(e)).
This region extends approximately from $\langle n\rangle=1.1$ to
$\langle n\rangle=1.45$ (see Fig.~\ref{PhaseDiagram-Away}) and
corresponds to a crossover from an $120^{\circ}$-AF to a ferromagnetic
(FM) phase. 
In this region, no peak structure has been detected in the spin susceptibility;
 these states are paramagnetic.
A further increase of the doping level results in a peak in the spin
susceptibility at $\mathbf{Q}=\mathbf{\Gamma}$, as shown in
Fig.~\ref{Sus-Peak}(f), indicating the formation of FM
correlations. 
The divergence of $\chi_{\Omega_{m}=0}^{spin}(\mathbf{\Gamma})$ represents the onset
of a stable FM phase. 

At the hole doped side, however, the destruction of
$\chi^{spin}_{\Omega_{m}=0}(\mathbf{K})$ as the hole concentration is
increased, occurs  much slower. 
We find a large region of $\chi^{spin}_{\Omega_{m}=0}(\mathbf{K})$
with peak structure, which spreads from $\langle n\rangle\sim1.0$ to
$\langle n\rangle\sim0.55$. 
This is in a sharp contrast to the electron doped side, where
$\chi^{spin}_{\Omega_{m}=0}(\mathbf{K})$ is quickly destroyed and
absent for $\langle n \rangle$ above $\langle n\rangle\sim1.1$.
In accordance with the suppression of $\chi_{\Omega_{m}=0}^{spin}$ at
$\mathbf{K}$, the amplitude of
$\chi_{\Omega_{m}=0}^{spin}(\mathbf{M})$ gradually increases.  
This indicates that  scattering processes with magnetic wave
vector $\mathbf{Q}=\mathbf{M}$ have larger and larger contributions, while
those with $\mathbf{Q}=\mathbf{K}$ become less important. 
Increasing the hole doping further,
$\chi_{\Omega_{m}=0}^{spin}(\mathbf{M})$ starts to peak at around $\langle
n\rangle\sim0.5$. 
This behavior corresponds to AF states of the collinear type (CAF).

The observation of  CAF in an isotropic system on a non-bipartite
lattice is rather unexpected. While it has been established that spin
anisotropy in Heisenberg models on the  triangular lattice favors
CAF~\cite{PhysRevLett.98.077205}, it usually requires higher order
hopping processes in an isotropic triangular 
lattice to bring about a CAF phase~\cite{PhysRevB.83.041104}.
In this work, we provide another alternative way of obtaining
CAF in {\it isotropic} triangular lattice, namely doping the system with
holes. The present case does, however, differ from the previously
discussed cases~\cite{PhysRevLett.98.077205,PhysRevB.83.041104} as no
long-range magnetic order is observed here.  
Another essential difference is that hole doping turns the system
metallic while the system is insulating in the previous cases.

\begin{figure}[htbp]
\centering
\includegraphics[width=\linewidth]{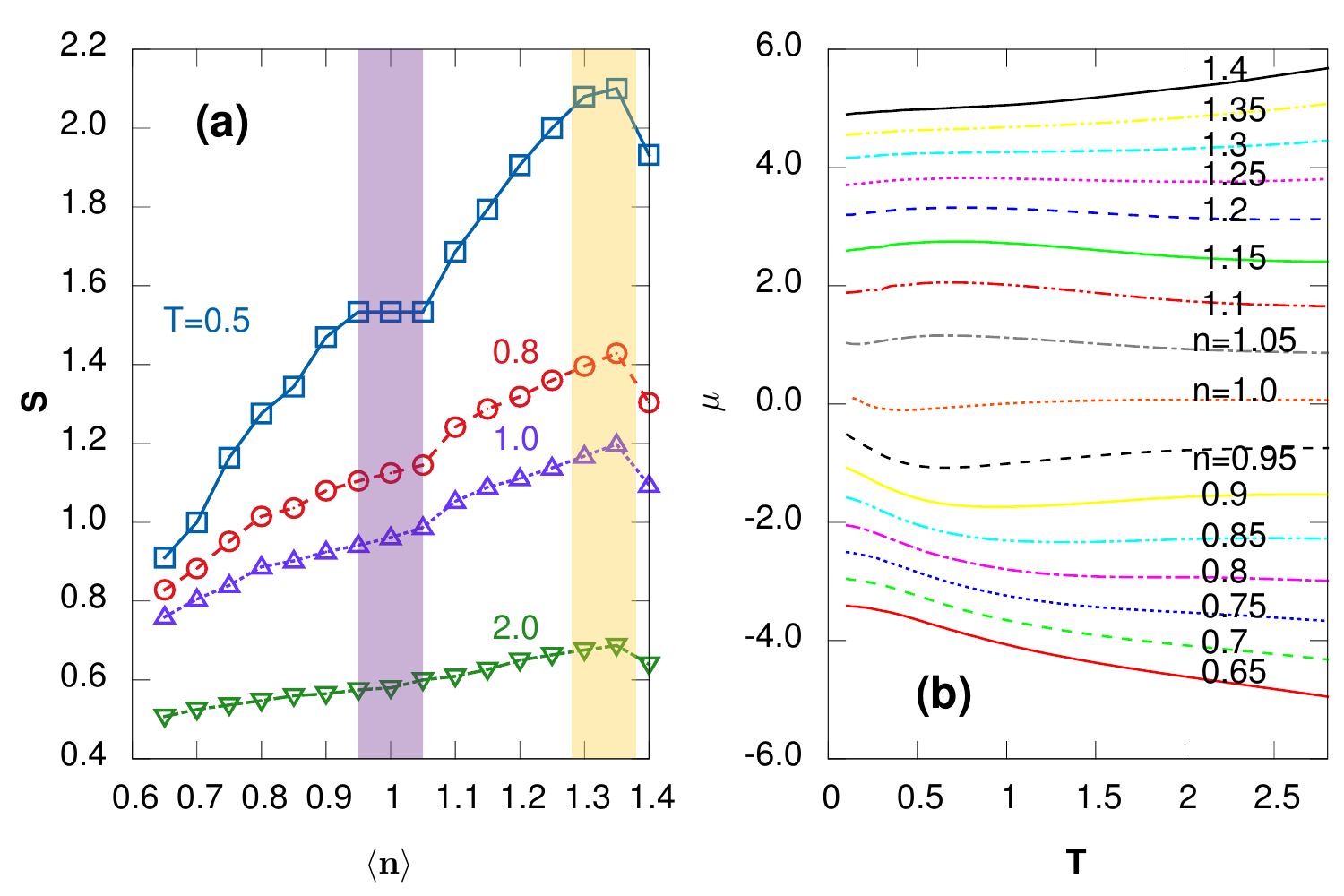}
\caption{(a) The doping dependence of the entropy for four different
temperatures with $U=W$. The two color filled area show the entropy
plateau at half-filling and maximization at $\langle n\rangle\sim
1.35$.  
(b) Temperature dependence of the chemical potential
  for a range of filling levels. From bottom to top, the curves correspond to
  $n=0.65$ to $n=1.4$ with $0.05$ as interval.} 
\label{mu-S}
\end{figure}

The entropy $S$ of the doped triangular system is shown in
Fig.~\ref{mu-S}(a). 
We determine the entropy by using the same procedure as for the
half-filled case.  
The calculation has been performed with $U$ fixed at $U=9$, which is
only slightly below $U_c^{MIT}$ where the system enters into a
magnetically ordered Mott insulating state. 
Thus,  spin and charge fluctuations are enhanced  and contribute
significantly to $S$. 
In Fig.~\ref{mu-S}(b), we show the chemical
potential $\mu$ vs. $T$ at various levels of fillings.
$S$ and $\mu$ are related via 
 the Maxwell relation
$\left({\partial S}/{\partial n}\right)_{T, U} =
-\left({\partial\mu}/{\partial T}\right)_{U, n}$.
The chemical potential shows opposite behavior (decreasing versus
increasing) at hole- and electron-doped sides (see
Fig.~\ref{mu-S}(b)). 
Correspondingly, the entropy displays different filling dependencies
with respect to electrons and holes. 
As already can be read off from Fig.~\ref{PhaseDiagram-Away}, the strong
AF correlations are destroyed quickly when doping the half-filled
system with electrons but persist to a large level of  hole doping.  
This is reflected in the behavior of $S$, which on the hole-doped side
is smaller than  on the electron-doped one.
We note that the Hubbard model on a square
lattice displays a similar behavior, but with the opposite electron
and hole dependence~\cite{PhysRevB.67.092509}. 
This implies certain similarities between the square and
triangular lattice. 
$S$ shows a plateau around half-filling, which relates to the nearly
constant $\mu$ for all temperatures at half-filling ($\langle
n\rangle=1$).

Motivated by the fact that $S$ in the Hubbard model on the square
lattice becomes maximal near optimal doping, {\it i.e.} $\langle
n\rangle\approx 0.85$, where the SC transition temperature is
peaked~\cite{PhysRevB.80.140505, doi:10.1080/000187300243381}. 
we locate the filling on the triangular lattice that maximizes $S$.
Interestingly, the thus obtained 'optimal' filling for the triangular system,
$\langle n\rangle\sim1.35$ coincides well with the optimal filling found in
Na$_{x}$CoO$_{2}\cdot$$1.3$H$_{2}$O~\cite{2003Nature.424.527,
  PhysRevLett.92.247001, PhysRevB.68.104508}. 
The maximization of the entropy reflects the strong competition
of the localized spin degrees of freedom (around half-filling) with
the charge degrees of freedom (here at large electron doping).
Enhanced entropy commonly occurs in the vicinity of quantum critical points.
In line with what is seen in the  cuprates~\cite{PhysRevLett.102.206407}
we, therefore, speculate that the ground state of the ITH model may
have a quantum critical point at this filling. We will return to this
issue in future work.

Our results of the filling dependence of $S$ allow us to
estimate the Seebeck coefficient via Kelvin's formula,
$S_{kelvin}=1/q_{e}(\partial S/\partial n)_{T,
  U}$~\cite{PhysRevB.82.195105}. 
We find that the thermopower is large (negative) at $\langle
n\rangle\sim 1.5$ but rather smaller at $\langle n\rangle\sim1.35$.
This nicely explains the different doping behaviors of the
thermopower experimentally found in
Na$_{x}$CoO$_{2}\cdot$$1.3$H$_{2}$O~\cite{PhysRevLett.92.247001,
  PhysRevB.79.075105}. 
When $\langle n\rangle$ is larger than $1.3$, $S$ starts
to decrease with a large negative slope, which coincides with the
large thermopower at $\langle n\rangle\sim 1.5$ in
Na$_{x}$CoO$_{2}\cdot$$1.3$H$_{2}$O. 

In conclusion, we studied the isotropic triangular Hubbard model using
the dual fermion approach to systematically incorporate non-local
correlations beyond the DMFT. 
By varying temperature, Coulomb interaction, and chemical potential, we
find that this model gives rise to a very rich phase diagram, which recovers
all the phases experimentally resolved in organic salts and
the recently much-studied layered cobaltate
Na$_{x}$CoO$_{2}\cdot$$1.3$H$_{2}$O (we do not explicitly calculate
the SC phase). 
Moreover, based on the behavior of the entropy, we find that the
isotropic triangular Hubbard model displays certain similarities to
the Hubbard model on square lattice for cuprates.  
Specifically, we find that the entropy is maximal at the optimal
doping for superconductivity of Na$_{x}$CoO$_{2}\cdot$$1.3$H$_{2}$O, 
and the experimentally observed different thermopower behaviors can be
nicely explained by the doping dependence of the entropy through the
Kelvin formula. 

{\it Acknowledgments:} We acknowledge the stimulating
post (at Nov. 09, 2011) of Ross H. McKenzie in his blog ``CONDENSED
CONCEPTS"~\cite{Condensed_Concepts}, which inspired our discussion on
the chemical potential and the entropy vs. filling.
We thank F.~Assaad and M.~Katsnelson for fruitful discussions.
G.L. and W.H. are financially supported by DPG Grant Unit
FOR1162.
One of us (W.H.) acknowledges the hospitality of the KITP at the
university of California in Santa Barbara, where part of this work was
completed and supported by the NSF under Grant No. NSF PHY11-25915.
A.N.R. acknowledges the RFFI grant 11-02-01443, Dynasty foundation and
S.K. the support under NSF Grant No.\ PHYS-1066293 and the hospitality
of the Aspen Center for Physics.

\bibliographystyle{apsrev4-1}
\bibliography{ref}

\onecolumngrid
\appendix

\section*{SUPPLEMENTAL MATERIAL}

\begin{figure}[htbp]
\centering
\includegraphics[width=0.8\textwidth]{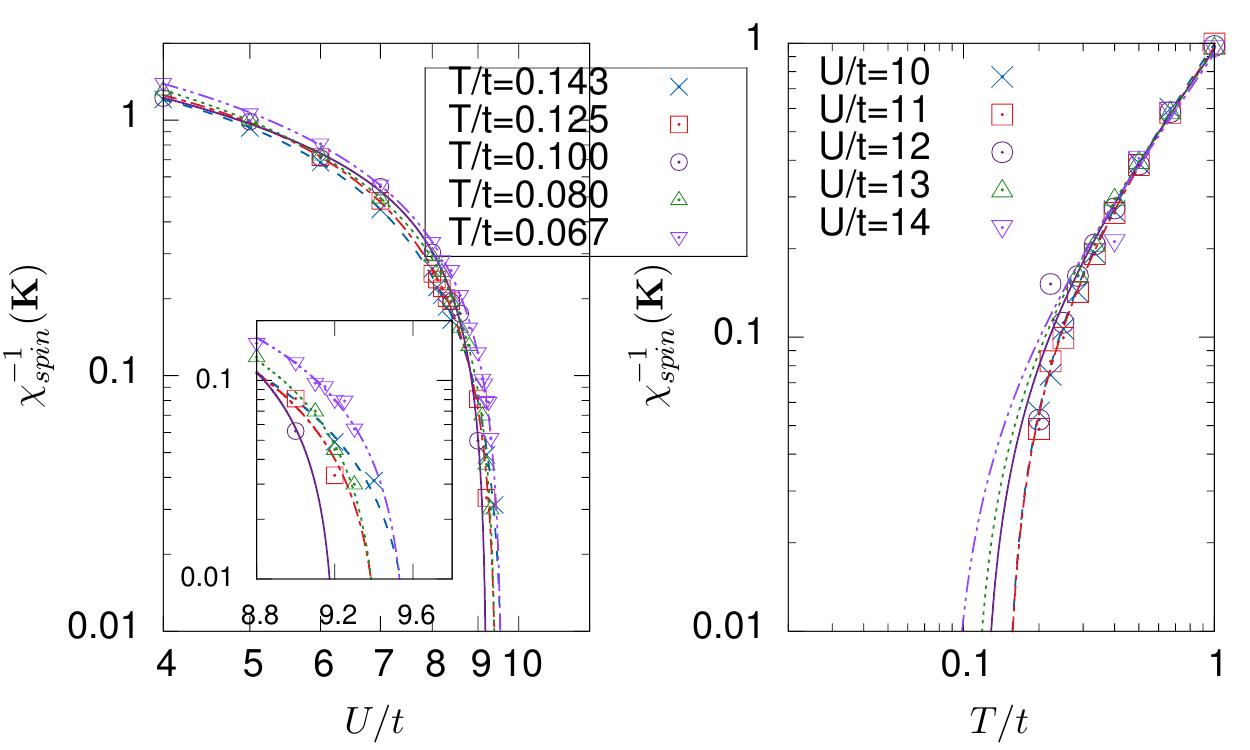}
\caption{Polynomial extrapolation of the inverse spin susceptibility as
  functions of interaction $U$ (left panel) or temperature $T$ (right
  panel).}
\label{Sup}
\end{figure}

The magnetic transition boundary is determined by extrapolating the spin
susceptibility $\chi_{spin}(Q, \Omega_{m}=0)$ with polynomials.  
For $\beta t\ge7$, we look for the divergence of $\chi_{spin}^{-1}(Q,
\Omega_{m}=0)$ at given temperature $T$ as a function of interaction
$U$, as shown in the left panel of Fig.~\ref{Sup}. 
For $U/t\ge10$ we extrapolate $\chi_{spin}^{-1}(Q, \Omega_{m}=0)$ for
each given interaction $U$ as a function of temperature $T$, see right
panel of Fig.~\ref{Sup}. 

\end{document}